\documentstyle[prl,aps,epsf,multicol]{revtex}
\begin{document}
\title{Quantum Sticking, Scattering and Transmission of $^4$He atoms
from Superfluid $^4$He Surfaces}
\author{C. E. Campbell$^{1,3,*}$, E. Krotscheck$^{1,2}$,
and M. Saarela$^{1,4,\dagger}$}
\address{$^1$Institut f\"ur Theoretische Physik, Johannes Kepler
Universit\"at Linz, A-4040 Linz, Austria}
\address{$^2$Department of Physics, Texas A\&M University,
College Station, TX 77843}
\address{$^3$School of Physics and Astronomy, University of Minnesota,
Minneapolis, Mn 55455}
\address{$^4$Department of Physical Sciences/Theoretical Physics,
University of Oulu, SF-90570 Oulu, Finland}
\date{\today}
\maketitle
\begin{abstract}

We develop a microscopic theory of the scattering, transmission, and
sticking of $^4$He atoms impinging on a superfluid $^4$He slab at near normal
incidence, and inelastic neutron scattering from the slab.  The theory includes
coupling between different modes and allows for inelastic processes.  We find a
number of essential aspects that must be observed in a physically
meaningful and
reliable theory of atom transmission and scattering; all
are connected with multiparticle scattering, particularly the
possibility of energy loss. These processes are (a) the coupling
to low-lying (surface) excitations (ripplons/third sound) which is
manifested in a finite
imaginary part of the self energy, and (b) the reduction of the
strength of the excitation in the maxon/roton region.
\end{abstract}
\begin{multicols}{2}
\bigskip
\narrowtext
\firstfigfalse

The dynamics of liquid $^4$He films and the bulk fluid near its free surface
continues to be of considerable interest. Experimental information is available
about the scattering of helium atoms from helium surfaces and
films\cite{Edwards75,EdwardsFatourosScattTh,Nayak83,%
SwansonEdwardsScatt4He,WyattReflectivity2}; from the dynamics of localized
excitations within the fluid, including excitation scattering from the surface
and quantum
evaporation\cite{WyattReflectivity2,WyattReflectivity1,WyattSpectrum,WyattRough}
;
and from inelastic neutron scattering at grazing angles from adsorbed
films\cite{LauterPhonon,LauterPRL,LauterJLTP,filmexpt}. Moreover, information
about the condensate fraction in helium may be obtainable directly from elastic
transmission of $^4$He atoms through superfluid $^4$He slabs\cite{HalleyDrops}.
The fact that helium slabs have been made in the laboratory \cite{WilliamsSlab}
leads to the prospect that the dynamic probes previously applied to adsorbed
films and bulk surfaces of helium can now be applied to slabs. The
slabs should produce simpler and thus easier to interpret results than the
adsorbed films.

We report here on the results of a manifestly microscopic
theoretical analysis of the dynamics of such a slab at zero
temperature. We find that atom scattering processes are
dominated by multi-particle events, particularly the coupling to
ripplons. This is in qualitative agreement with the conclusions of
Edwards and
collaborators\cite{Edwards75,EdwardsFatourosScattTh,Nayak83,%
SwansonEdwardsScatt4He} in their results for the related helium atom
scattering from the free surface of bulk helium.  Our
results not only provide insight into the transmission and sticking of a
helium beam that was not previously available from theory or experiment, it
also provides detailed predictions for the anticipated experiments on
slabs.

The theoretical method used here has been successfully applied to
the bulk\cite{Chuckphonon,Mikkou2}, to adsorbed
films\cite{filmdyn,filmexc} and to droplets\cite{HNCdrops}; in the latter
two it was successful in making detailed and correct predictions about
surface states (ripplons and third sound) which also play a very important
role in the slab geometry.  The theory is adapted here to the slab
geometry, including scattering states.  We outline our theory and
present some illuminating results waves for at non-normal incidence
which also provide insights into quantum evaporation.

\begin{figure}[b]
\epsfxsize=3truein
\centerline{\epsffile{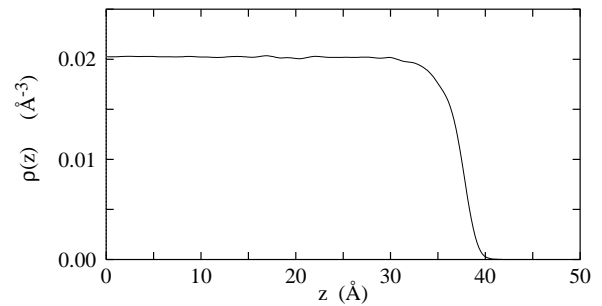}}
\bigskip
\caption{The figure shows the density profile of the $^4$He slab used
 here. The profile is symmetric around
$z=0$.\label{slabprofile}}
\end{figure}

We consider the ground state of a slab of superfluid $^4$He of
particle number $1.5~{\rm\AA}^{-2}$, corresponding to a thickness of
approximately $80~{\rm\AA}$, and the dynamic structure function that
would be measured for neutron momentum loss perpendicular to the slab.
The density profile of this slab in the ground state is shown in
Fig. \ref{slabprofile}.
The relevant excited states may be written as
\begin{equation}
\Psi_\lambda({\bf r}_1, \dots, {\bf r}_N) =
F_\lambda({\bf r}_1, \dots, {\bf r}_N) \Psi_0({\bf r}_1, \dots, {\bf r}_N),
\end{equation}
where $\Psi_0$ is the ground state of the slab or an appropriately
optimized representation of the ground state, $F_\lambda$ is a
complex {\it excitation operator,\/} and $\lambda$
represents the quantum numbers for these excited states.
The excitation energy for these states is given by
\begin{equation}
\epsilon_\lambda=
{{\left\langle\Psi_\lambda \vert H - E_0 \vert \Psi_\lambda\right\rangle}\over
{\left\langle \Psi_\lambda \vert\Psi_\lambda \right\rangle}}=
{\sum_{i=1}^N {{\left\langle\Psi_{0} \left\vert
{\hbar^2\over{2m}}\left|\nabla_iF_{\lambda}\right|^2\right\vert
\Psi_{0}\right\rangle}}\over{{\left\langle \Psi_\lambda \vert\Psi_\lambda
\right\rangle}}}.
\end{equation}
The functions $F_\lambda$ are solutions of an effective
Schr\"odinger equation obtained by functionally
minimizing this excitation energy with respect to $F$ \cite{Feynman},
by using correlated
\noindent
basis function (CBF) perturbation
theory\cite{Chuckphonon}, or by extremizing the
action\cite{Mikkou2,filmdyn}.

\begin{figure}[b]
\epsfxsize=3truein
\centerline{\epsffile{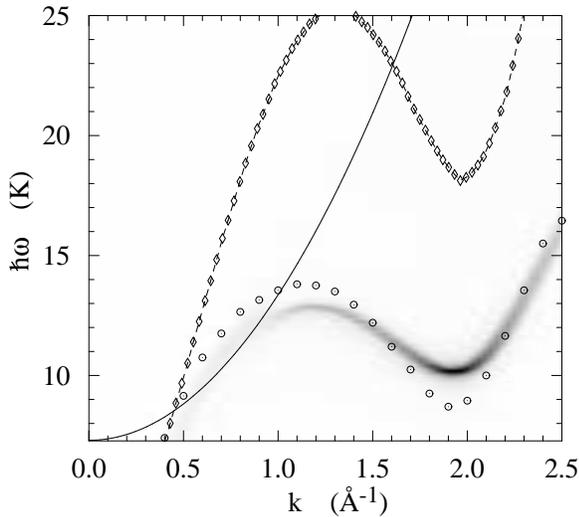}}
\bigskip
\caption{The figure shows a grey-scale map of the dynamic structure
function $S(k_\perp,\omega)$ for momenta $k_\perp$ perpendicular to
the slab, and energies above the evaporation barrier $\hbar\omega >
-\mu$. Also shown are (a) the experimental phonon-roton
spectrum\protect\cite{CowleyWoods}, (circles), the Feynman spectrum
$\hbar^2k^2/[2mS(k)]$ (dashed line with diamonds) and the kinetic
energy of the incoming particle, $-\mu + \hbar^2 k^2/2m$ (solid line).
\label{skwperp}}
\end{figure}

It is well-known from the theory of the phonon-roton spectrum of bulk liquid
$^4$He that these low excited states are quantitatively accounted for by
a wavefunction of this type containing one- and two-body
terms in $F_\lambda$\cite{Chuckphonon,Mikkou2,filmdyn,filmexc}:
\begin{equation}
F = \sum_{i=1}^Nf_1({\bf r}_i) + \sum_{i<j=1}^Nf_2({\bf r}_i, {\bf r}_j).
\label{Fequation}
\end{equation}
Retaining only $f_1$ in the bulk case produces the familiar Bijl-Feynman
spectrum $\epsilon_{\bf k}=\hbar^2k^2/[2mS(k)]$ where $k$ is the
wavenumber for the bulk excitation and $S(k)$ is the zero temperature x-ray
structure factor. This excitation energy is quantitatively
correct at long wavelengths and qualitatively correct in the
maxon-roton regime ({\it cf.\/} Fig. \ref{skwperp}). Including
$f_2$ is sufficient to correct most of the residual disagreement with
experiment\cite{Chuckphonon,Mikkou2,filmexc}.

Our work reported herein is a further adaptation of the above
procedure for  studying
transmission, reflection and sticking of incident particles.  This
is achieved by solving our equations with the boundary
condition that there is an incoming particle beam of specified energy and unit
incoming flux, and an outgoing particle beam.
Far from the
slab the particle is a plane wave with wave number determined by the
energy and the direction of propagation.

For simplicity, in this letter we focus on the elastic transmission and
reflection of states with normal incidence. Thus we may describe some of
our results in terms of reflection and transmission amplitudes
$R$ and
$T$, respectively. At first glance we
appear to be describing a one-dimensional quantum mechanical scattering problem
with the helium slab serving as a well or barrier,  but the actual
situation is far richer: Since this scattering ``well'' is composed of helium
atoms, this is a generically non-local problem when viewed at the one-body
level.  The Bose exchange symmetry of the incoming atoms with those in the slab
requires a full symmetrization, as one sees in the
summations in
Eq. (\ref{Fequation}).  Moreover, the well is dynamic: the incoming particle
may produce excited states, corresponding to inelastic processes, which may
result in the capture of the particle and/or the emission of particles in
states other than the elastic channel; this includes multi-excitation
channels in which the individual states carry momentum parallel to the
surface.  Of particular importance are the effects of surface
states, where the incoming particle may stick; and the structure of the
surface, which can disperse the incoming particle into a number of states
of quasi-momentum differing significantly from the momentum of the incoming
particle.

Since the states that we are exploring will also be excited by
inelastic neutron scattering, it is useful to first consider the
dynamic structure factor $S({\bf k}, \omega)$ that would be
measured by neutrons scattered with momentum change
$\hbar{\bf k}$.  This is obtained theoretically by using
linear response theory to obtain the density-density response function
$\chi({\bf k}, \omega)$ together with the relation $S({\bf
k}, \omega)=-{\cal I}m~\chi({\bf k}, \omega)/\pi$.  In the case
of the bulk superfluid at low temperatures,
$S(k,\omega)$ has a very sharp spectrum that maps out the bulk phonon-roton
excitation energy as well as broad multi-excitation strength at higher
energies. Neutron scattering from adsorbed helium films produces a similar
phonon-roton structure when studied as a function of the parallel momentum
transfer $\hbar{\bf k}_\parallel$ in grazing angle
scattering\cite{filmexc}.  However the layered structure of these films
broadens this phonon-roton structure and produces surface and layer modes
which are also detected in the neutron scattering, but which complicate the
analysis \cite{filmexc} and interpretation of data.  There is no
layering in the slab ({\it cf.\/} Fig. \ref{slabprofile}); thus the
broadening of
the phonon-roton structure is significantly reduced for grazing angle
scattering. Nevertheless, surface modes are still present and would be observed
similarly to the adsorbed film system.
However, if there is significant momentum transfer perpendicular  to the
surface,
one would expect significant surface effects on
$S({\bf k}; \omega)$ particularly for relatively thin slabs.
Nevertheless it can be seen from Fig. \ref{skwperp} that the calculated
structure in $S$ for our $80~{\rm\AA}$ slab
has  substantial strength in the vicinity of the bulk phonon-roton
spectrum for perpendicular momentum transfer ${\bf k}_\perp$,
though it is noticeably broadened and weakened. The fact that $S({\bf k}_\perp,
\omega)$ is effectively parametrizable in terms of
${\bf k}_\perp$ should not be interpreted to indicate that this is a good
quantum number. Nevertheless it is clear from the figure that it is
useful to use ${\bf k}_\perp$ to approximately classify these modes.

To examine the propagation of a helium atom normally incident at
energy $\hbar \omega$, we first exhibit the wave equation satisfied by the
one-body part of the excitation function $F$.  Defining the auxiliary function
$\psi({\bf r}) =f_1({\bf r})/\sqrt{\rho_1({\bf r})}$, the equation
for $\psi$ is:
\begin{eqnarray}
&&{{\hbar^2}\over{2m}}\left\{-\nabla^2 +
{\nabla^2 \sqrt{\rho_1({\bf r})}
\over\sqrt{\rho_1({\bf r})}}\right\}\psi({\bf r})
+ \int \Sigma ({\bf r},{\bf r}'; \omega)\psi({\bf r}')d^3 r'\nonumber\\
&& = \hbar \omega \left\{\psi({\bf r}) + \int d^3 r'
\left[g_2({\bf r},{\bf r}')-1\right]
\sqrt{\rho_1({\bf r})\rho_1({\bf r}')}\psi({\bf r}')\right\}\nonumber\\
\label{scatteq}
\end{eqnarray}
where $\rho_1$ and $g_2$ are the density and pair distribution of the
ground state, and $\Sigma ({\bf r},{\bf r}'; \omega)$ is
the self-energy. This equation has the appearance of a one-body
Schr\"odinger equation with a non-local, non-hermitean ``optical
potential'', which has its origin in the fact that this is a many-body
system.
The derivation, and the approximations we use to
calculate the self energy may be found in Ref. \onlinecite{filmexc}. In our
approximation, it has the form
\begin{equation}
\Sigma({\bf r},{\bf r'};\omega) = -{1\over 2}\sum_{mn}
        {V_{mn}({\bf r}) V_{mn}({\bf r'})
	\over\hbar(\omega_m+\omega_n-\omega)-i\eta}
\label{Sigmaeq}
\end{equation}
where the $V_{mn}({\bf r})$ are three-phonon vertex functions derived
in Ref. \onlinecite{filmexc},
and $\omega_m$ are the excitation energies of the background; note
that the state sums go over both phonon and ripplon type excitations
and include, in particular, all parallel momenta. This feature is
manifested in the energy denominator which can cause the self-energy
to be complex.

The excited states consist of bound and continuum states. The scattering states
of a $^4$He atom from the slab are continuum states. Thus equation
(\ref{scatteq}) is solved subject to the boundary conditions
\begin{equation}
\psi({\bf r}) = \left\{\begin{array}{r@{\quad{\rm for}\quad}l}
e^{ikz} + R e^{-ikz} & z\rightarrow-\infty\\
T e^{ikz} & z\rightarrow \phantom{-}\infty
\end{array}\right.
\end{equation}
where $k$ is the positive root of $~\omega = \hbar k^2/2m$.

We have carried out calculations including both the full, complex
self-energy as well as the simpler, Feynman approximation. The latter is
equivalent to setting $f_2({\bf r}, {\bf r}')=0$ in the definition of the
excitation factor $F$, which gives $\Sigma({\bf r}, {\bf r}';\omega)=0$ in
equation (\ref{scatteq}), reducing the equation to the one used by
Edwards and Fatouros\cite{EdwardsFatourosScattTh} to
describe scattering from the surface of the bulk liquid.  At that level, the
stationary scattering states have the unphysical property that the elastic
single-particle flux is conserved, {\it i.e.\/} $|R|^2 + |T|^2 = 1.$

\begin{figure}[b]
\epsfxsize=3truein
\centerline{\epsffile{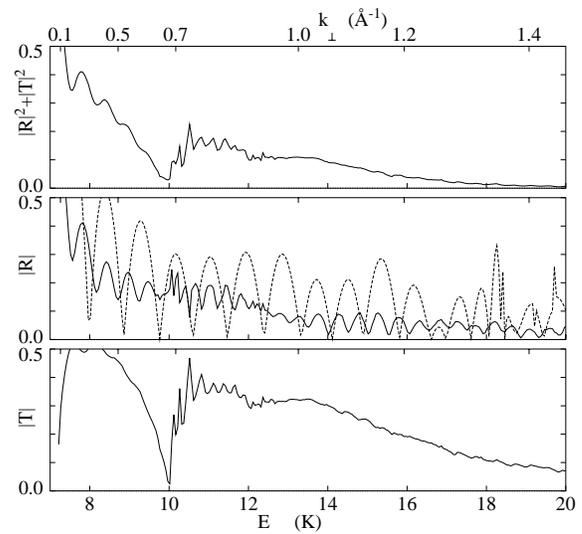}}
\bigskip
\caption{The absolute value of the transmission and reflection
coefficients, $T$ (lowest figure), $R$ (middle figure) and the
intensity loss $|R|^2 + |T|^2$ (top figure) are shown as obtained from
the CBF calculation (solid lines). We also show the Feynman
approximation for $|R|$ (middle figure, dashed line); note that $|R|^2
+ |T|^2 = 1$ in that approximation.\label{r2t2}}
\end{figure}

Our results for $|R|$ and $|T|$ are summarized in Fig. \ref{r2t2}. It is
seen in
the middle panel of Fig. \ref{r2t2} that the reflection coefficient $|R|$
undergoes oscillations similar to those seen in scattering from wells
and barriers in one-dimensional one-body quantum mechanics. The
details are of course different due to the fact that the effective
dispersion relation inside the slab is quite different from the
free-particle spectrum, as can be seen in Fig. \ref{skwperp}.

The self-energy has dramatic effects on the results:

The {\it real part\/} of the self-energy comes from virtual processes
that dress the Feynman states by allowing for fluctuations of the
short-ranged structure of the system. The consequence is a significant
improvement of the single excitation energies, as is seen in
Fig. \ref{skwperp}.

The {\it imaginary\/} part of the self-energy comes from the existence
of multi-excitation states with total energy equal to the single
excitation state.  Thus a single particle impinging on the surface has
channels for decaying into these multi-excitation states. (In our
nomenclature, a multi-excitation state is one in which the excitation
function $F$ is primarily a product of single excitation $F$ factors.
Thus, e.g., a two excitation state would be characterized by a
dominant term in equation (\ref{Fequation}) of the form $f_{2,\omega}({\bf
r},{\bf r}')=f_{1,{\omega_a}}({\bf r})f_{1,{\omega_b}}({\bf
r}')+f_{1,{\omega_b}}({\bf r})f_{1,{\omega_b}}({\bf r}')$, where
$\omega = \omega_a + \omega_b +O(N^{-1}$).)  One consequence of this
is that the incoming atom can ``stick'' in the slab by decaying to two
or more bound states.  Similarly it can decay into bound states and an
emitted particle of lower energy.  This leads to a significant
reduction of $\left\vert R\right\vert^2+\left\vert T \right\vert^2 <
1$ as is seen in the top panel of Fig. \ref{r2t2}.  The difference $1
-\left[\left\vert R\right\vert^2+\left\vert T \right\vert^2\right]$ is
a measure of the atomic sticking plus real inelastic scattering
and transmission.

The change in the transmission coefficient is the most dramatic effect
of allowing for decay processes: Hermitean approximations for the
self-energy all lead to the feature $|R|^2+|T|^2=1$. In this aspect,
our predictions differ dramatically from those of the Feynman
approximation as well as a more recent
attempt\cite{DalfPitStringEvapLett} to study quantum evaporation
within time-dependent density functional theory. Our conclusions
agree, however, with those of Edwards and
Fatouros\cite{EdwardsFatourosScattTh} when damping of reasonable
magnitude is included.

Another interesting feature seen in Fig. \ref{r2t2} is that the
transmission coefficient has notable structure at and above the roton
energy which is largely missing from the reflection coefficient. This
structure is in part due to the high density of states of
the roton, and to the fact that the phonon-roton spectrum is non-monotonic
in this region, giving rise to degeneracies for energies between the
roton and the maxon.  Moreover the absence of translational
invariance in the $z$ direction results in the incoming plane-wave
hybridizing with the degenerate states in the slab.  This is a surface
effect, and would continue to exist for very thick films and for the
surface of the bulk.

An examination of the wavefunctions and the structure of
the self-energy shows that most of the sticking occurs in the surface
of incidence, with a further reduction of amplitude at the back
surface.  By turning off the surface state contributions to the
imaginary part of the self-energy (ripplons/third sound and other
states localized in the surface) we have seen that the main
contribution to this surface sticking comes from these surface states.

In conclusion, our adaptation of the microscopic theory of excited states in
inhomogeneous liquid $^4$He to describe quantum sticking, scattering
and transmission of $^4$He atoms gives a clear picture of the
many-body physics of the interaction of a beam of $^4$He atoms with a
liquid helium surface.  However the existence of a second surface, at
only approximately $80~{\rm\AA}$ behind the first in the present work, makes a
direct comparison to scattering from the free surface of bulk liquid $^4$He
ambiguous.  A significant fraction of the scattering from the slab occurs
at the
second surface, contributing some signal to the reflection by back
propagation.  Some of the sticking also happens at the second surface.  The
remaining elastic transmission into the vacuum
can also be viewed as quantum evaporation, as can be seen by
examining the propagation of localized wavepackets, wherein
excitations in the interior of the slab, produced by the incoming
wavepacket at the first surface, propagate to the second surface where
helium atoms are evaporated.

The main approximation in this work is that only decay into two-excitation
states is included.  Opening other inelastic
channels would further reduce the amplitude of the elastically scattered and
transmitted atoms.

This work was supported in part by NASA grant NAGW3324 and the
Minnesota Supercomputer Institute [CEC], by NSF grant DMR-9509743 and
the Austrian Science Fund under
project P11098-PHY [EK], and the Academy of Finland [MS]. An honorary
Fulbright Grant during this work is gratefully acknowledged by
CEC.
%
%

%


%

\end{multicols}
\end{document}